\documentclass[11pt]{article}
\usepackage[applemac]{inputenc}
\usepackage{graphicx}
\usepackage{amssymb}
\usepackage{lscape}
\begin{document}
\title{THE ALHAMBRA-SURVEY\\For a systematic Study of Cosmic Evolution}

\small
\author{Mariano Moles$^{(1)}$ (PI), Emilio Alfaro$^{(1)}$,
Narciso Ben\'{\i}tez$^{(1)}$, Tom Broadhurst$^{(2)}$,
\and
Francisco J. Castander$^{(3)}$, Jordi Cepa$^{(4, 5)}$,
Miguel Cervi\~no$^{(1)}$,
\and
Alberto Fern\'andez-Soto$^{(6)}$, Rosa M. Gonz\'alez Delgado$^{(1)}$,
Leopoldo Infante$^{(7)}$,
\and
Alfonso L\'opez Aguerri$^{(4)}$, Isabel M\'arquez$^{(1)}$,
Vicent J. Mart\'{\i}nez$^{(6)}$,
\and
Josefa Masegosa$^{(1)}$, Ascensi\'on del Olmo$^{(1)}$,
Jaime Perea$^{(1)}$,
\and
Francisco Prada$^{(1)}$, Jos\'e Mar\'{\i}a Quintana$^{(1)}$, Sebasti\'an S\'anchez$^{8}$}
\normalsize
\vspace{-2truecm}
\maketitle
\vspace{-1truecm}
\begin{flushleft}
\begin{small}
(1) Instituto de Astrof\'{\i}sica de Andaluc\'{\i}a, CSIC, Granada,
Spain\\
(2) School of Physics and Astronomy, Tel Aviv University, Israel\\
(3) Institut d'Estudis Espacials de Catalunya, CSIC, Barcelona,
Spain \\
(4) Instituto de Astrof\'{\i}sica de Canarias, La Laguna, Spain\\
(5) Departamento de Astrof\'{\i}sica, Facultad de F\'{\i}sica, Universidad de la Laguna, Spain\\
(6) Observatori Astron\`omic de la Universitat de Val\`encia,
Val\`encia, Spain \\
(7) Departamento de Astronom\'{\i}a, Pontificia Universidad
Cat\'olica, Santiago de Chile, Chile\\
(8) Centro Astron\'omico Hispano-Alem\'an, CAHA, Almer\'{\i}a, Spain \end{small}
\end{flushleft}
\begin{flushright}

{\it\bf La realidad aplaca lo imaginario como el r\'\i o la bruma} \\
{\sl Maim\'onides}
\end{flushright}

\normalsize
\abstract

The ALHAMBRA-Survey is a project to gather the data necessary to sample a cosmologically significant fraction of the Universe with enough precision to follow the evolution of its content and properties with $z$, a kind of {\sl Cosmic Tomography}. It is defined as a large area, 4$\square^{\circ}$, photometric survey with 20 contiguous, equal width, medium band filters covering from 3500 \AA~ to 9700 \AA, plus the standard JHK$_s$ near-infrared bands. The photometric system in the optical was optimized to get (for a fixed amount of total observing time) the maximum number of objects with accurate classification and redshift and to be sensitive to relatively faint emission features in the spectrum.  We expect to be able to detect emission features down to EW = 35 \AA, for S/N $\approx$ 30.

The observations will be carried out with the 3.5m telescope in Calar Alto (Spain) using the new wide field cameras in the optical (LAICA) and in the NIR (OMEGA-2000). We intend to reach the limit AB = 25 mag (for an unresolved object, S/N=5) in all the optical filters from the bluest to 8300~\AA, and from 24.7 to 23.4 for the remainder. The expected limit in the NIR is fixed at K$_s$ = 20 mag, H = 21 mag, J = 22 mag.

The homogeneous and contiguous spectral coverage will result in several hundred thousand objects with accurate SED identification and z-values. The accuracy of the survey will allow us to study, among others, the large scale structure evolution with $z$, the identification of clusters of galaxies (with membership assessment for a significant fraction of the galaxies), the identification of families of objects, and other detailed studies, without the need for any further follow-up. Indeed, it will provide exciting targets
for large telescopes, the GTC in particular. Given its area and spectral coverage and its depth, apart from those main goals, the ALHAMBRA-Survey will also produce valuable data for galactic studies.


\section{Introduction: Global scientific aim and opportunity}

One of the major issues in Cosmology is {\sl\bf Cosmic Evolution}. The central issue is to disentangle genuine cosmic evolution from physical variance at a given redshift and the details of the metric, what has been a permanent challenge for Physical Cosmology. To approach the question of Cosmic Evolution meaningfully it is therefore necessary to sample large physical volumes even at low redshifts, to capture not only representative average properties but also their variance.

This implies a combination of wide area and depth, with a continuous spectral coverage to avoid having complex selection functions depending on z and the nature of the objects. Indeed, the quest for precision implies a good enough spectral resolution and photometric accuracy.

Some of the previous requirements have been met by several photometric surveys already finished or still under way, but not all of them at the same time. Thus, the SDSS has produced a rather shallow sampling of a huge area, whereas deeper surveys have sampled the distant and/or faint Universe in
rather small areas.

Up to now, the largest surveys were photometric, ensuring a complete spectral coverage with broad-band filters. The resulting precision in $z$ obtained via photometric redshift techniques ($\sim 0.1$ in $\Delta z/(1+z)$, at best) and in Spectral Energy Distribution (SED) determination are correspondingly rough. At the other extreme in spectral resolution, spectroscopic surveys cannot go as deep as the photometric surveys, reaching only $I\approx 24$ with the use of large telescopes. The covered fields are necessarily small and cannot cope with the complex variety of objects in the Universe.

To get the optimum compromise between large area and depth, good spectral resolution and coverage, we have defined the \textbf{A}dvanced \textbf{L}arge, \textbf{H}omogeneous \textbf{A}rea \textbf{M}edium \textbf{B}and \textbf{R}edshift \textbf{A}stronomical, \textbf{ALHAMBRA}-Survey. It is a photometric survey primarily intended for cosmic evolution studies. We propose to cover a large-area with 20 contiguous, equal width, medium band optical filters from 3500 \AA\ to 9700 \AA, plus the three standard broad bands, JHK$_s$, in the NIR. Thus, it is placed halfway in between the traditional imaging and spectroscopic surveys.

It will make possible the study of many different astronomical problems in a self-contained way. By design, the ALHAMBRA-Survey will provide precise ($\Delta z < 0.015(1+z)$) photometric redshifts and SED classification for $\geq 300,000$ galaxies and AGNs. Thanks to the unbiased nature of this survey (i.e. neither designed to detect a given class of objects nor to be precise in fixed windows only), important problems other than Cosmic Evolution can be addressed. These include the study of stellar populations in the galactic halo, the search for very cold stars and blue stragglers, and the possible detection of debris from galactic satellites in the Milky Way halo. Moreover, the large surveyed area and the ability to finely discriminate between different spectral energy distributions will permit the serendipitous detection of objects that could be classified as {\em exotic} or {\em rare}. This broad category includes very high redshift galaxies ($\approx$ 2500 objects at $z>5$, with $\Delta z <0.01$, expected from scaled HDF observations) and QSOs.

The opportunity to undertake such a project was prompted by the new access to the Calar Alto Observatory (Almer\'{\i}a, Spain), now shared on equal basis by Germany and Spain. The ALHAMBRA-Survey received the support of a large fraction of the Spanish astronomical community and was allocated
Guaranteed Spanish Time.

%
%
\section{The project implementation}

The idea to use photometric filter information to determine the redshift of faint sources was first proposed by Baum (1962), and later re-launched by Loh \& Spillar (1986) and Koo (1986) as a {\sl poor person} machine to get redshifts.

The ALHAMBRA-Survey is a multi-narrowband survey with complete spectral coverage in the visual range. We have defined the filter system to have a complete, homogeneous spectral coverage in the visual domain, and added the NIR survey to complement the information about the detected objects and to improve the $z$ and SED determination, in the case of relatively large photometric errors (see below) or for particular classes of objects. It has also been adapted to the instrumental capabilities available now in Calar Alto. The filter system has been designed to optimize the output in terms of z and SED determination accuracy. The immediate precedent was the survey proposed by Hickson {\sl et al} (1994), with a filter system very different from what we proposed here, as we show below.

The main scientific goal of the ALHAMBRA-Survey is to provide the community with a set of data appropriate for the systematic study of Cosmic Evolution. The hypothesis of homogeneity and isotropy implies the existence of maximally symmetric subspaces and the existence of a 1-to-1 relation between redshift and time. This is a model-independent prediction, prior to any consideration about the value of the cosmological parameters. Precisely, we intend to materialize a {\sl foliation of the space-time}, producing narrow slices in the $z$-direction whereas the spatial sections are large enough to be cosmologically representative, what could be called {\bf\sl Cosmic Tomography}

\subsection{The ALHAMBRA-Survey filter system}

We have optimized the number of filters a) to get accurate SED and $z$ determination for the largest possible number of objects, for a given total amount of observing time, and b) to be sensitive to relatively faint emission lines. Given the performance of the instruments to be used, the total exposure time was fixed to 100 ksec. As for the filters, we have analyzed four cases: constant or logarithmic ($\Delta\lambda\propto(1+\lambda)$) width, with either minimal or 50\% overlap. In all cases the filters continuously cover the whole optical interval from 3500~\AA\ to 9700~\AA, with almost square transmission efficiency.

To understand the true efficiency of the system we generated a mock catalogue with 13,000 galaxies upon the magnitude, redshift and spectral type distribution of the galaxies in the Hubble Deep Field (Fern\'andez-Soto {\sl et al}, 1999) and the photometric redshifts calculated with the BPZ software (Ben\'{\i}tez 2000; templates as in Ben\'{\i}tez {\sl et al} 2004). Since the accuracy of the input photometric redshifts is $\approx0.06(1+z)$, we perturb them by a similar, randomly distributed amount to produce a more realistic redshift distribution. We generate magnitudes in each of the filter systems above with realistic photometric noise added with the estimated performance of the site + 3.5m telescope + LAICA cameras. We distribute the exposures trying to reach constant $S/N$ per filter, but with two constraints: the minimal exposure time per filter is, for practical reasons, at least $2,500$s, and we do not expose more than twice this time in a homogeneous exposure distribution, to avoid spending all our time on the less efficient filters.
%
\begin{figure}
\includegraphics[angle=0,width=13cm,height=9cm]{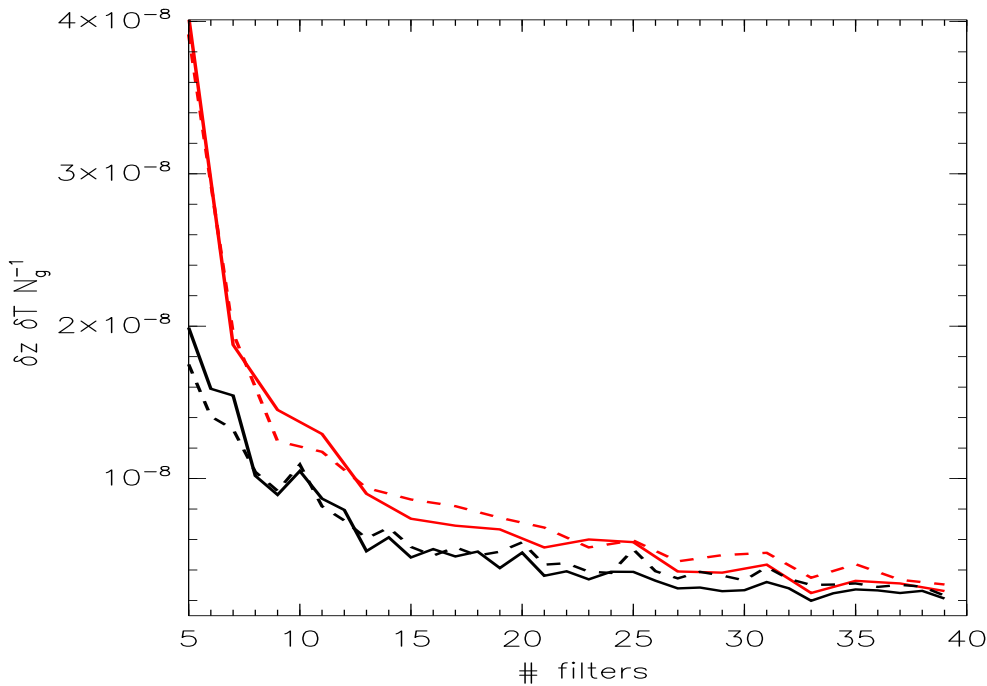}
\caption{{\small Product of the normalized rms in photometric redshift by the rms in spectral type, divided by the number of objects with $Odds=1$}. Solid (dashed) lines are for constant (logarithmic) width, with (red) and without (black) overlap}
\label{Nz}
\end{figure}
%
The results are presented in Ben\'{\i}tez {\sl et al}, these proceedings. Here we give only the main result, illustrated in Figure~\ref{Nz} (here only the information from the optical filters is considered, for the impact of including the NIR filter information, see below). It is clear that the constant width, non-overlapping filter system is the best performer for any fixed number of filters. We also find that the improvement in the number-weighted precision of the survey is slow after $n_{f}= 15$ filters. Therefore, from that point of view, the conclusion would be to use 15 non overlapping filters, of 410 \AA\ width to cover the whole spectral range.

Our second requirement was the possibility to detect relatively faint emission lines. The median value of the H$\beta$ line of HII galaxies amounts to 35 \AA, whereas 85\% of the same galaxies have EW([OII]) $\geq$ 35 \AA\ (Terlevich {\sl et al} 1991). Therefore, detecting lines with observed EW lines of 35 \AA\ or stronger would allow to find a substantial fraction of this, or other similar, family of galaxies. Now, an emission line of equivalent width EW in one of the filters would be detected at the n-$\sigma$ level provided that

$$ EW \geq n\sqrt{2}\sigma\times W_F$$

\noindent
where W$_F$ is the filter width and $\sigma$ the error in each of the measured magnitudes. Detection at the 3$\sigma$ level for photometric errors of 0.03 mag would then imply $ EW \geq 0.127 W_F$. It is clear that to detect a substantial fraction of the emission line objects similar to HII-galaxies the filters should be narrower than the 410 \AA\ wide filters resulting form the previous simulations. To detect a line with EW = 35 \AA, S/N = 30, at the 3$\sigma$ level, filters $\approx$300\AA~ wide are needed.

Therefore, the ALHAMBRA optical photometric system was designed to include 20 contiguous, medium-band, FWHM = 310~\AA, square-like shaped filters with marginal overlapping in $\lambda$, covering the complete optical range from 3500 to 9700 ~\AA. With this configuration it is possible to accurately determine the SED and $z$ even for faint objects and to detect rather faint emission lines. With this configuration the H$\beta$ line of a typical HII-like galaxy could be detected till $z \approx 1$ for rest frame EW values over 20 \AA, whereas the [OII] line would be detectable till $z \approx 1.6$, with rest frame EW of over 14 \AA.

We also notice that a galaxy like IZw18 could be detected till z = 0.1

\begin{figure}
\begin{center}
\includegraphics[angle=0,width=15cm,clip=true]{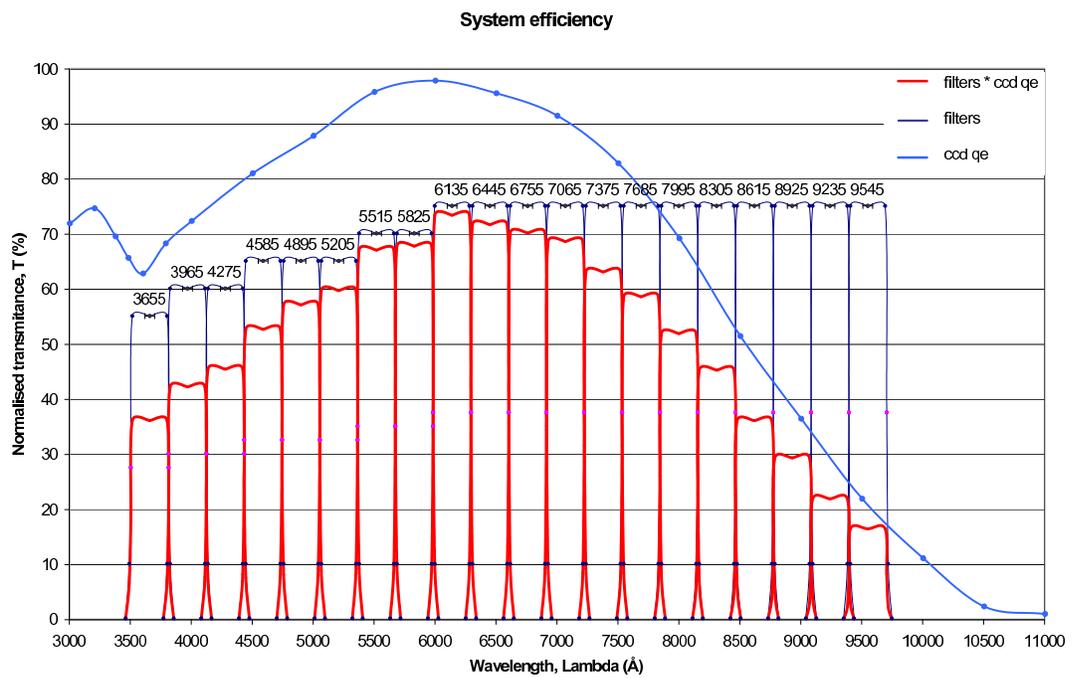}
\caption{{\small Calculated transmission curves of the ALHAMBRA-system taking into account the atmospheric transmission and the CCDs quantum efficiency, for the minimum filter transmission guaranteed by the manufacturer}}
\label{eficiencia}
\end{center}
\end{figure}

In the Figure~\ref{eficiencia} we have plotted the transmission curves of the filters as offered by BARR, together with the quantum efficiency of the detectors and the final response of the system.

The survey will also include NIR deep observations through the standard broad band filters, JHK$_s$.

\section{Field selection}

\subsection{The total covered area}

Although the Universe is in principle homogeneous and isotropic on large scales, astronomical objects are clustered on the sky on different scales. The clustering signature contains a wealth of information about the structure formation process. A survey wanting to study clustering needs to probe as many scales as possible. In particular, searching contiguous areas is important to cover smoothly the smallest scales where the signal is stronger and obtain an optimally-shaped window function. Measuring a
population of a certain volume density is a Poissonian process with an associated variance. One would obtain different densities of the same population when measuring in different places. The variance in those measures is dictated by the volume density of the population under study, the volume searched and the clustering of the population. In order to beat down this sample (or cosmic) variance one needs to sample independent volumes. So there should be a balance between probing contiguous area and
independent areas. On the technical side, the geometry of LAICA, the 3.5m Calar Alto telescope instrument with which the ALHAMBRA survey is devised to be carried out in the optical, imposes a minimum contiguous area patch of $1^{\circ} \times 0.25 ^{\circ}$.

The relative error in any counting statistical measure will scale as $\sqrt{\frac{N_{st}}{N_f\;\Delta\;z}}$, where $N_{st}$ is the number of spectral types that need to be resolved, $N_f$ is the number of fields and $\Delta\;z$ the redshift interval to be resolved. Assuming the SDSS local luminosity function with a plausible evolutionary parameterizations that yield a redshift distribution similar to the one obtained from the HDF at our magnitude limit, we estimate that we will get relative errors of $\sim
0.015\sqrt{\frac{N_{st}}{N_f\;(\Delta\;z/0.1)}}$ at redshift $z\sim 0.3$ where the SDSS starts to be unable to sample the galaxy population due to its relatively bright flux limit, relative errors of $\sim 0.010 \sqrt{\frac{N_{st}}{N_f\;(\Delta\;z/0.1)}}$ at redshift $z\sim0.8-1.0$; of $\sim 0.010 \sqrt{\frac{N_{st}}{N_f\;(\Delta\;z/0.1)}}$ at redshift $z\sim1.5$ and of $\sim 0.020 \sqrt{\frac{N_{st}}{N_f\;(\Delta\;z/0.1)}}$ at redshift $z\sim2.0$. So, if we decide that we want relative errors no larger than 2\% at $z\sim2.0$ with a redshift resolution of $\Delta\;z = 0.1$ and resolving 8 spectral types, we would need 8 square degrees. At any other lower redshifts our relative errors will be lower.

We first considered that a total are of 8 square degrees would be optimum given the general conditions and aims of the project. This would however represent a too big demand of photometric telescope time to be completed in a reasonable time, say no more than 3-4 years. Therefore, we have defined an observing strategy aiming at completing two strips, 1$^{\circ}\times$0.25$^{\circ}$ each, in each of the selected fields, to ensure a large enough covered area and a good sampling to cope with the cosmic variance.

\subsection{The selected fields}

To select the fields to be covered we have taken into account the following,
evident criteria:

\begin{enumerate}
\item {Low extinction}
\item {No (or few) known bright sources}
\item {High galactic latitude}
\item {Overlap with other surveys and/or other wavelengths}
\end{enumerate}
The selected fields are listed in Table~\ref{campos}.

\begin{table}
\caption{{\small The ALHAMBRA-Survey Selected Fields}}
\begin{center}
\label{campos}
\begin{tabular}{|l|c|c|c|c|c|c|c|}
\hline
Field name                      &  RA(J2000)  & DEC(J2000) & 100 $\mu$m   & E(B-V)  & l & b \\
\hline
\hline
ALHAMBRA-1                  & 00 29 46.0 & +05 25 30 & 0.72 & 0.017 & 113 & -57 \\
\hline
ALHAMBRA-2                  & 01 30 16.0 & +04 15 40 & 0.80 & 0.022 & 140 & -57  \\
\hline
ALHAMBRA-3/SDSS         & 09 16 20.0 & +46 02 20 & 0.67 & 0.015 & 174 & +44 \\
\hline
ALHAMBRA-4/COSMOS   & 10 00 00.0 & +02 05 11 & 0.90 & 0.018 & 236 & +42 \\
\hline
ALHAMBRA-5/HDF-N     & 12 35 00.0 & +61 57 00 & 0.60 & 0.011 & 125 & +55 \\
\hline
ALHAMBRA-6/GROTH    & 14 16 38.0 & +52 24 50 & 0.35 & 0.007 &  95 & +60 \\
\hline
ALHAMBRA-7/ELAIS-N1 & 16 12 10.0 & +54 30 15 & 0.27 & 0.005 &  84 & +45 \\
\hline
ALHAMBRA-8/SDSS        & 23 45 50.0 & +15 35 05 & 1.17 & 0.027 & 99 & -44 \\
\hline
\end{tabular}
\end{center}
\end{table}

\section{Global expectations}

\subsection{Sensitivity considerations. Instrument performances.
Calibration Strategy}

\begin{figure}
\begin{center}
\includegraphics[angle=-90,width=12cm]{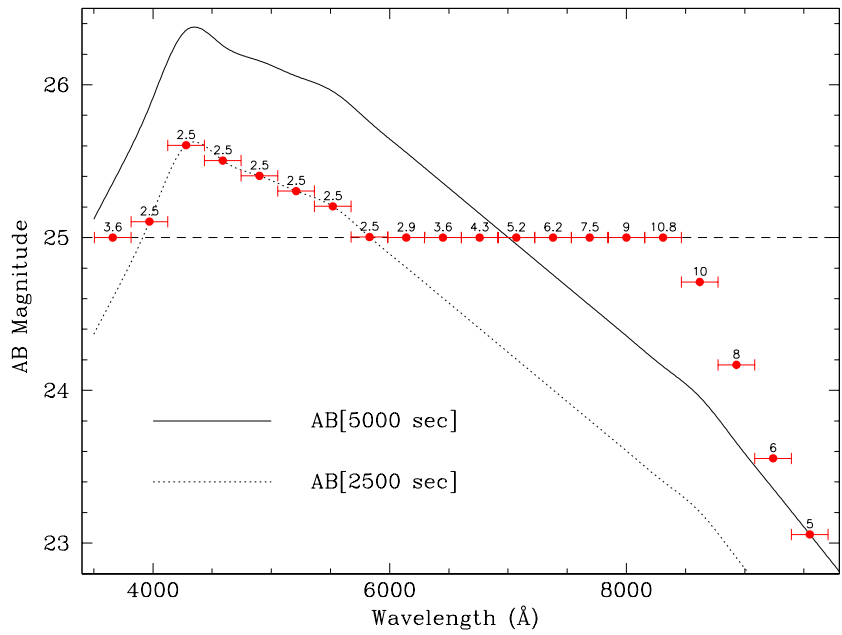}
\caption{{\small Total exposure time per filter in ksec. For the bluest
filters the time is fixed by logistic considerations rather than for
accuracy reasons}}
\label{expt}
\end{center}
\end{figure}

Taking into account the average extinction in Calar Alto and the performance
of the telescope and cameras, we have calculated the exposure time per
filter to reach the proposed limit. The results are plotted in
Figure~\ref{expt}. They were calculated for AM = 1.3, FWHM = 1.2. The
average exposure time per filter amounts to 5000s, for the reasons discussed
before. In the bluest filters the exposure time is fixed by the need to get
a minimum number, actually 5, of exposures to correct for cosmic and
transitory artifacts. In the reddest filters the exposure time is limited to
fit within the total exposure time allowed for a given pointing.

In the NIR the available information on the Omega2000 performances allows us
to estimate that we will be able to reach K$_s$ = 20, H = 21, J = 22 in
5000s per filter and pointing.

To calibrate the optical data we intend to determine the extinction
coefficients and calibrations points through observation of the appropriate
standard stars. To that end it is not necessary to obtain them for all the
20 ALHAMBRA filters. A significantly smaller number of filters would be
enough to determine the input parameters of the models shown to fit very
well the atmospheric extinction at Calar Alto (see Hopp \& Fern\'andez
2002). This question will be settled empirically.

To calibrate photometrically the data we need to establish a network of
standard stars covering a wide colour range, at least up to spectral type
M0. ALHAMBRA-20 being a new photometric system, we have to start defining
the zero points for the magnitudes. The core of the ALHAMBRA network of
standard stars (ANSS) is a set of stars with accurate spectrophotometric
calibration. We will use this set as primary calibrators at the telescope
and to define the secondary calibrators, chosen to be stars in the target
fields.

The magnitudes will be defined from the catalogue fluxes by:

$$ m = -2.5 \rm{log} \frac{\int_{F}f(\lambda)S_F(\lambda)d\lambda}
{\int S_F(\lambda)d\lambda} + Cte$$

This is the usual way to calibrate narrow-band images when the photometric
system is not previously defined, as for line filters. The SDSS Consortium
has also adopted that strategy to define their own photometric system
(Fukugita {\sl et al}, 1996; Smith {\sl et al}, 2002). We will set the ANSS
on the AB system,

$$AB_{\nu}=-2.5log f_{\nu}-48.574$$

\noindent where $f_{\nu}$ is the flux per unit frequency from an object
in erg s$^{-1}$ cm$^{-2}$ Hz$^{-1}$

The primary standard stars will be selected form the lists by Oke \& Gunn
(1983), Oke (1990), Massey \& Gronwall (1990) and Stone (1996), together
with the 4 fundamental calibrators adopted by the HST.

For practical reasons (optimize the use of the 3.5m telescope) but also to
control the accuracy and homogeneity of the photometric calibration, we have
considered to define secondary standards in every subfield corresponding to
a CCD in the LAICA imager, 15.4'x15.4' each. The plan is to select 3-5 stars
in each field and to obtain accurate, calibrated absolute spectrophotometry
of all of them. The observations will be carried out with the 1.5m OSN
telescope with the ALBIREO spectrograph. These secondary standards will be
observed under photometric conditions and the extinction and calibration
will be secured by observing primary standards.

The NIR observations are done in the standard system J,H,K$_s$. The
procedure is also standard, with observations of standard stars to determine
the night extinction and the zero points of the system. For similar reasons
as in the LAICA case, we have decided to have secondary standards in every
single field of $\Omega$2k. The observations will be done at the 1.5m CST at
the Iza\~na Observatory (Tenerife, Spain). An important aspect to control is
the differences (hopefully small) between the Calar Alto and Iza\~na
systems. The transformations will be determined from the observations of the
same (primary) stars in both sites.

\subsection{The number of objects with accurate SED and z determination}

\begin{figure}
\includegraphics[angle=0,width=12cm]{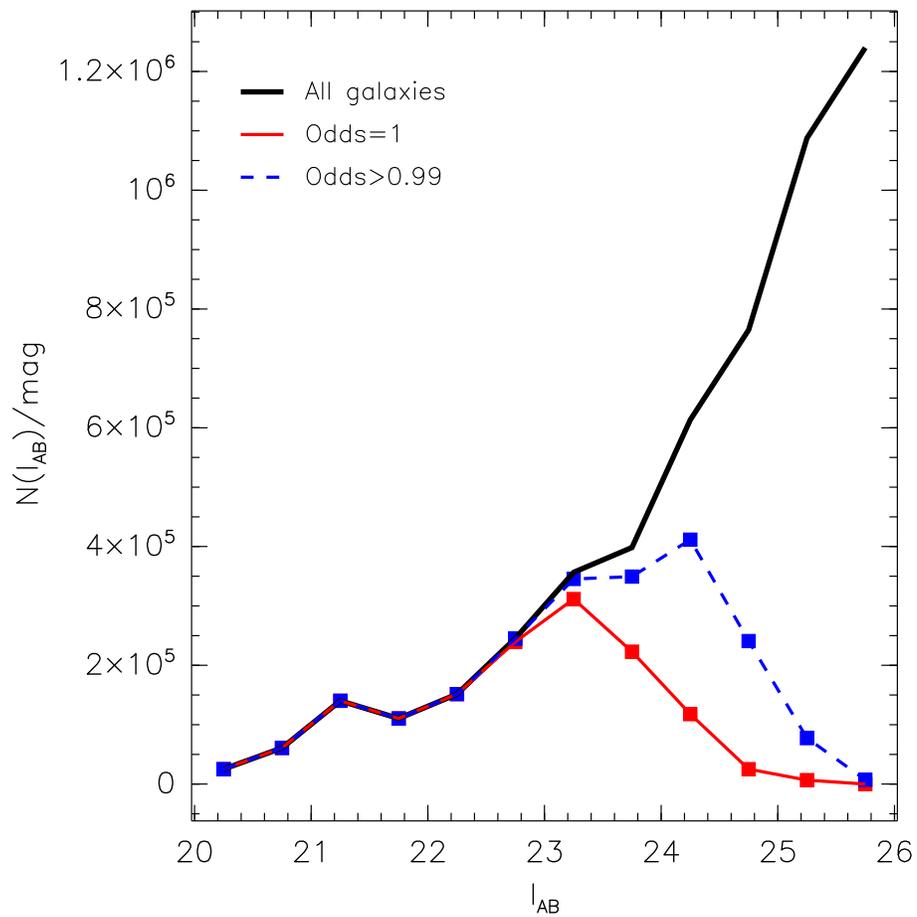}
\caption{{\small Number of galaxies with $Odds>0.99$ and $Odds=1$ as a
function of magnitude}}
\label{Nmag}
\end{figure}

In Figure~\ref{Nmag} it can be seen that we can obtain highly accurate,
$Odds=1,\Delta z/(1+z)\approx0.015$ redshifts for $\approx90\%$ of galaxies
with $I_{AB}<23.5$, a total over 300,000. If we relax the selection criteria
to $Odds>0.99$, we would then reach $90\%$ completeness at $I_{AB}=24$, with
a photo-z accuracy of $\Delta z/(1+z)\approx0.03$ (more than 500,000
galaxies). The results have been obtained for the simulations described
before. Let us point out that this is a minimum since we intend to analyze
the implementation and use new and more detailed and specific templates than
those used in the simulations, that could improve the quality of the
fittings and the final results.

\begin{figure}
\begin{center}
\includegraphics[angle=0,width=12cm]{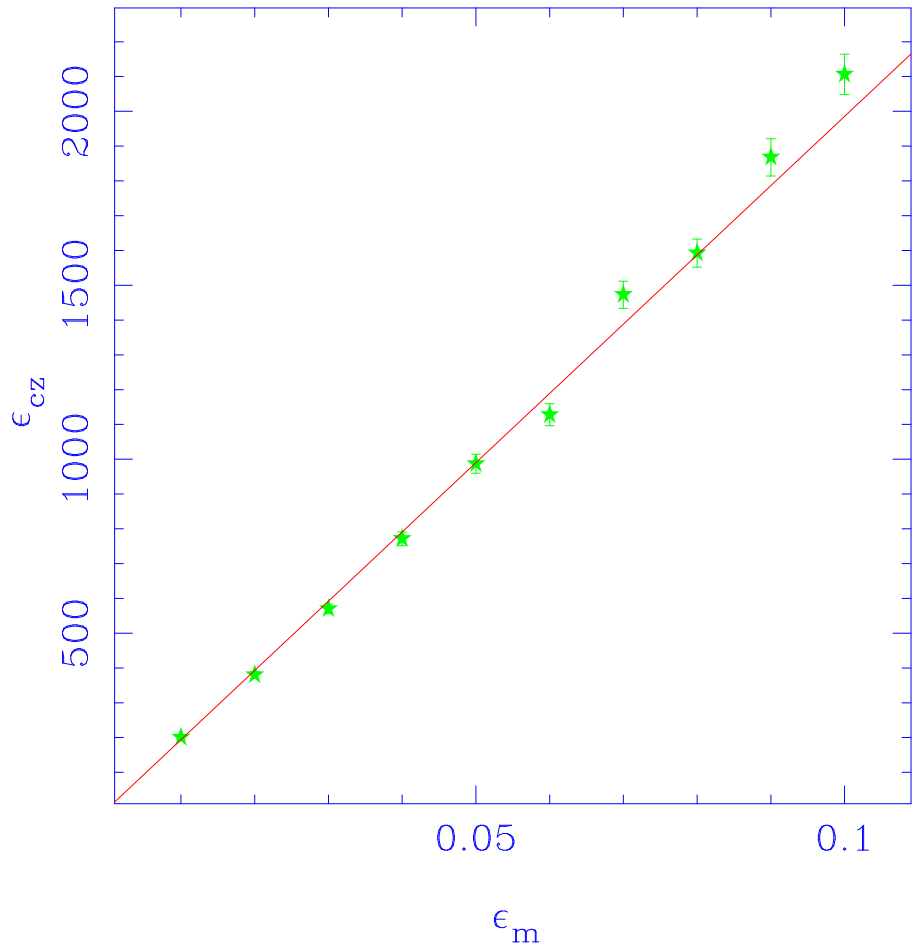}
\caption{{\small The errors in the redshift determination as a function
of the input photometric errors}}
\label{errors}
\end{center}
\end{figure}

Indeed, the final result critically depends on the photometric errors and
the adequacy of the templates. To test the first aspect we have built a mock
catalogue of $~5\times10^{5}$ galaxies, with z between 0 and 1.5, from the
sets of templates given by Kinney (E, S0, SA, Sc), Bica (Starburst),
Calzetti (Starburst) and a QSO. White noise with different amplitudes was
added to the objects in order to calibrate the effect of photometric
errors. Then, the redshift of each target was determined using best fitting
criteria by the initial templates. The results are shown in
Figure~\ref{errors}, where we plot the errors resulting in z as a function
of the input photometric error (error bars were obtained by bootstrap) in
magnitudes.

\begin{figure}
\includegraphics[angle=0,width=14cm]{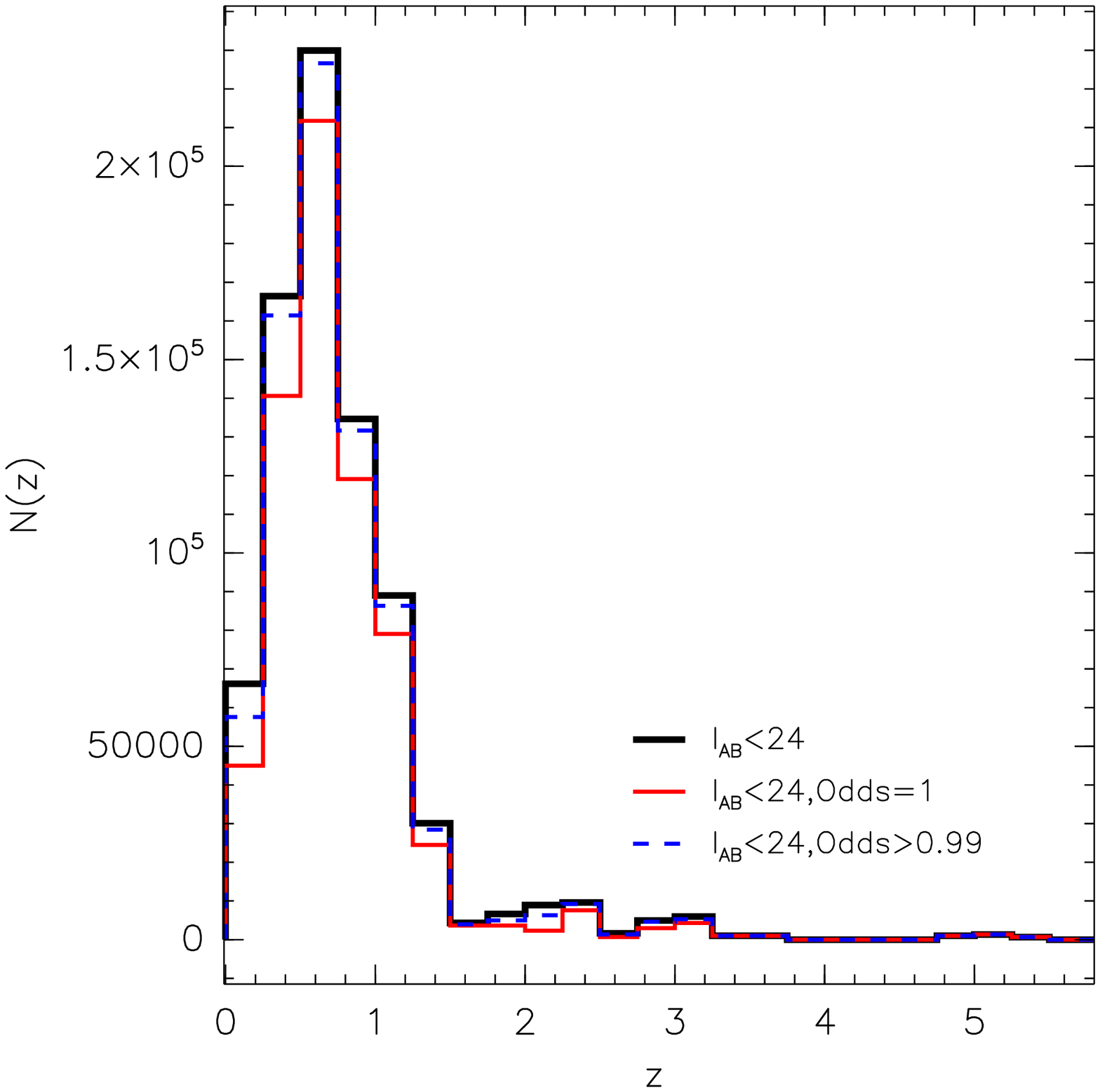}
\caption{{\small Number of galaxies with $I_{AB}<24$, $Odds>0.99$
and $Odds=1$ as a function of redshift}}
\label{Nred}
\end{figure}

Finally, we show in Figure~\ref{Nred} the efficiency as a function of
redshift. As expected we can see than the efficiency of the photo-z is very
low in the $1.5<z<3.$ interval, but otherwise is rather homogeneous. Note
that most high-z objects will have good redshift measurements.

Inclusion of the NIR information can significantly improve the z
determination in some cases. In particular, as has been pointed out by many
authors using photometric redshift techniques in deep surveys, the use of
NIR filters can help break the degeneracy between low-redshift ($z \approx
0.5$) and high-redshift ($z \approx 3$) galaxies. The reason behind this
degeneracy is the confusion between the Balmer and Lyman breaks, which are
the most salient features of the respective spectral energy
distributions. In absence of any infrared information, it is not possible to
tell the slope of the rest-frame red end of the spectrum, the range that can
in fact tell the difference between both families of objects.

\begin{figure}
\begin{center}
\includegraphics[angle=0,width=7cm]{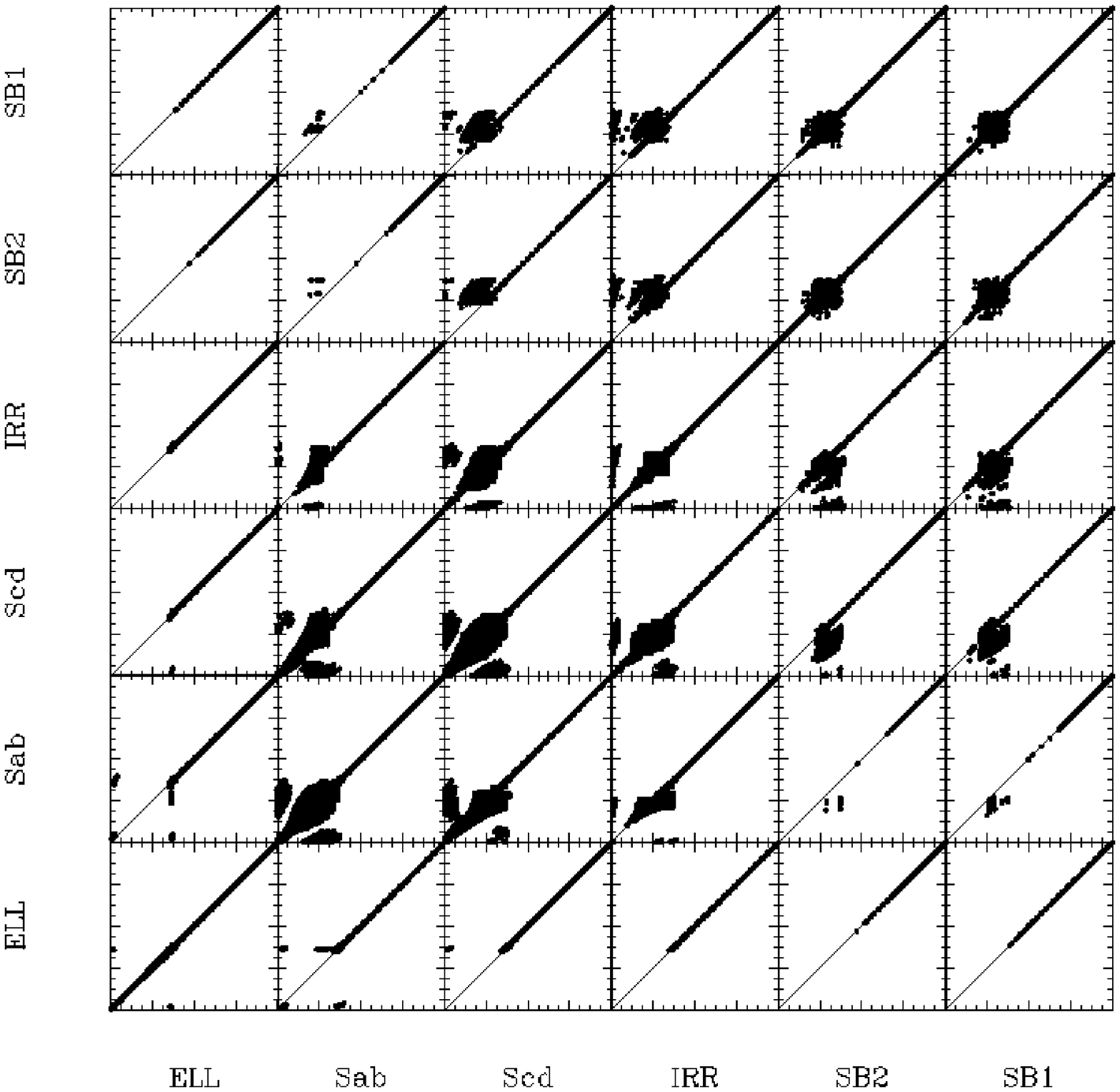}
\includegraphics[angle=0,width=7cm]{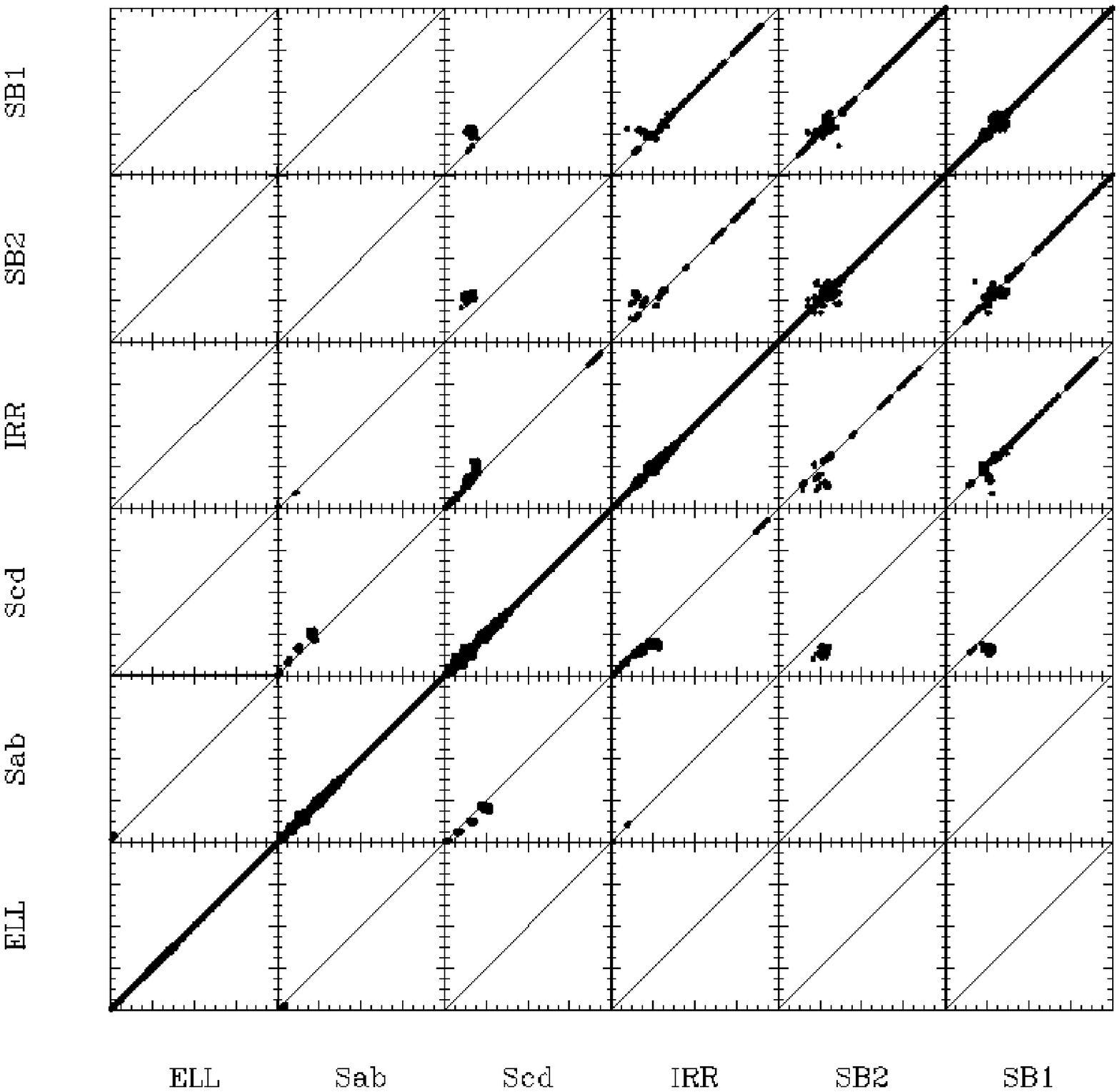}
\caption{{\small Theoretical degeneracies in type and redshift expected for
galaxies in the ALHAMBRA survey measured to an accuracy of 0.2 magnitudes in
all filters. The left panel shows the case where no infrared information is
available, and the right panel corresponds to the case where infrared
information is included. Each of the $6 \times 6$ subpanels corresponds to a
$z_1$ vs $z_2$ diagram with redshifts ranging from 0 to 8.}}
\label{infrared}
\end{center}
\end{figure}

Figure~\ref{infrared} explicitly shows this effect. Each panel shows the
theoretical degeneracies between the different spectral types and redshifts
(all redshift axes range from $z=0$ to $z=8$). It can be seen how (for a
typical case) the presence of NIR data elliminates most of the degeneracies
between low and high redshift, and sharply separates the elliptical, Sab,
and Scd galaxies from the rest and from each other, leaving only some
residual degeneracy between the three bluest types. Of course, we should
never forget that the infrared information also adds greatly to the
information content of the survey, via the more direct relation existing
between the galactic mass and the infrared luminosity.

\section{Conclusions}

The ALHAMBRA survey aims at filling a yet empty niche in astronomical
surveys, halfway between relatively shallow, wide-area spectroscopic
surveys, and deep, narrow-area photometric surveys. We intend to observe
a large area (4 square degrees) divided in eight separate sub-fields using a
specially designed set of 20 mid-band, non-overlapping filters covering the
whole visible range from 3700 \AA\ to 9700 \AA, plus the standard $JHK_s$
near infrared filters.

The survey has been designed having in mind the use of photometric redshift
techniques as the basic analysis tool. We have carried on detailed
simulations based on available deep catalogues, and estimate that we can
measure high-quality redshifts and accurate spectral types for more than
300,000 galaxies with $\Delta z/(1+z)\approx0.015$, and for more than half a million galaxies down to $I_{AB} \approx 24$, with redshift accuracy $\Delta z/(1+z)\approx0.03$. Approximately 2000 of these galaxies will be at $z>5$.

The main objective of our survey is the study of cosmological evolution,
under the many facets it can offer. We will study the evolution of the large
scale structure, the evolution of the populations of different cosmic
objects, and the processes leading to galaxy formation, evolution, and
differentiation. The unbiased nature of the survey will also allow for the
study of many different kinds of objects, ranging from emission-line
galaxies to the diverse types of AGNs, and stars in our own Galaxy.

At the present stage of the project we have started data collection, with
the first runs having taken place in August 2004, December 2004 and February 2005. The next run scheduled for May 2005. The core team (formed by the authors
of this paper) has already designed the data analysis routines, and the
first version of the data analysis pipeline will be operative by mid
2005. We intend to complete the survey data acquisition after six semesters and offer the survey products to the community two
years after the acquisition of the last data.

\end{document}